\documentclass[fleqn,usenatbib, onecolumn]{mnras}

\usepackage{mathptmx}

\usepackage[T1]{fontenc}

\usepackage{graphicx}	
\usepackage{amsmath}	
\usepackage{amssymb}	
\usepackage{upgreek}
\usepackage{xspace}
\usepackage{siunitx}
\usepackage{physics}
\usepackage{multirow}
\usepackage[version=4]{mhchem}
\usepackage{pdflscape}
\usepackage{xcolor}
\usepackage{changes}
    \setaddedmarkup{\blue{#1}}
    \setdeletedmarkup{\red{\sout{#1}}}

\newcommand{\ASigma}{{$\mathrm{A}\,{}^2\Sigma^+$}\xspace}
\newcommand{\BPi}{{$\mathrm{B}\,{}^2\Pi$}\xspace}
\newcommand{\CPi}{{$\mathrm{C}\,{}^2\Pi$}\xspace}
\newcommand{\XPi}{{$\mathrm{X}\,{}^2\Pi$}\xspace}

\newcommand{\DSigma}{{$\mathrm{D}\,{}^2\Sigma^+$}\xspace}

\newcommand{\duo}{\textsc{Duo}\xspace}
\newcommand{\exocross}{\textsc{ExoCross}\xspace}
\newcommand{\noname}{\texttt{NOname}\xspace}
\newcommand{\marvel}{\textsc{MARVEL}\xspace}
\newcommand{\XABC}{\texttt{XABC}\xspace}

\newcommand{\Ldoubling}{{$\varLambda$\,-\,doubling}\xspace}
\newcommand{\abinitio}{\textit{ab~initio}\xspace}
\newcommand{\NO}{\ce{^{14}N^{16}O}\xspace}

\newcommand{\red}[1]{{\color{red} #1}}
\newcommand{\blue}[1]{{\color{blue} #1}}
\definecolor{mygreen}{HTML}{00C290}

\title[ExoMol XLII: NO UV line list]{ExoMol molecular line lists -- XLII: Rovibronic molecular line list for the low-lying states of NO}

\author[Q. Qu et al.]{
Qianwei Qu,
Sergei N. Yurchenko,
and Jonathan Tennyson\thanks{E-mail: j.tennyson@ucl.ac.uk}
\\
Department of Physics and Astronomy, University College London, Gower Street, London WC1E 6BT, UK
}

\date{\red{The manuscript has been accepted for publication in
 \textit{Monthly Notices of the Royal Astronomical Society}.}}

\pubyear{2021}

\begin{document}
\label{firstpage}
\pagerange{\pageref{firstpage}--\pageref{lastpage}}
\maketitle

\begin{abstract}
An accurate line list, called \texttt{XABC}, 
is computed for nitric oxide which
covers its pure rotational,
vibrational and rovibronic spectra. A mixture of
empirical and theoretical electronic transition dipole moments
are used for the final calculation of $^{14}\mathrm{N}^{16}\mathrm{O}$ 
rovibronic  $\mathrm{A}\,^2\Sigma^+$ -- $\mathrm{X}\,^2\Pi$, 
$\mathrm{B}\,^2\Pi$ -- $\mathrm{X}^2\Pi$ and 
$\mathrm{C}\,^2\Pi$ -- $\mathrm{X}\,^2\Pi$ 
which correspond to the $\upgamma$, $\upbeta$ and $\updelta$ band systems, respectively, as well as  minor improvements to transitions
within the $\mathrm{X}\,^2\Pi$ ground state.
The work is a major update of the ExoMol \texttt{NOname} line list.
It provides a high-accuracy NO ultraviolet line list
covering the complicated regions where the 
$\mathrm{B}\,^2\Pi$-$\mathrm{C}\,^2\Pi$ states interact.
\texttt{XABC} provides comprehensive data for
the lowest four doublet states of NO
in the region of $\lambda > 160 ~ \mathrm{nm}$ 
($\tilde{\nu} < 63~000~\mathrm{cm}^{-1}$)
for the analysis of atmospheric NO on Earth, Venus or Mars,  other astronomical observations
and  applications. The data is available via www.exomol.com. 
\end{abstract}




\section{Introduction}
    Nitric oxide (NO) is widely distributed in the universe.
    The molecule was observed in the atmosphere
    of Venus \citep{08GeCoSa.NO} and Mars \citep{08CoSaGe.NO}, where it is one of the emission sources of the UV nightglow \citep{90BoGeSt,05BeFrSe}.
    Gerin et al. detected transitions of NO in the dark clouds
    L134N \citep{92GeViPa.NO} and TMC1 \citep{93GeViCa.NO}.
    \citet{01HaApZi.NO} reported their analysis of transitions of \ce{N2O} to \ce{NO}
    in the core region of the Sagittarius B2
    and evaluated the N/O chemical network.
    The first detection of extragalactic NO by \citet{03MaMaMa.NO}
    helps us to understand the chemistry of galaxy NGC 253.
    NO has yet to be detected in the atmosphere of an exoplanet but is thought likely to be important in the atmospheres of rocky exoplanets \citep{20ChZhYo.NO}.
    The detection of NO in the astronomical objects relies on knowledge of the corresponding spectral lines of the molecule so accurate NO line list plays significant role in the processes.
    
    There are several available
    high-resolution line lists of the NO \XPi ground state.
    The HITRAN database by \citet{jt691} is widely used for
     investigations of the Earth's atmosphere and other room temperature studies.
    For higher temperature applications, the NO line list in the HITEMP database has been 
     recently updated \citep{jt763} based upon use of the ExoMol \noname line list \citep{jt686}.
    The CDMS database by \citet{CDMS} contains long-wavelength data including lines with
    hyperfine-structure;
    data on two vibrational bands of \NO are available on  the CDMS website. 
    \citet{jt686} published the \noname lines list
    as a part of the ExoMol database \citep{jt810}.
    \noname, available for six isotopologues,
    contains \num{21688} states and \num{2409810} transitions for $^{14}$N$^{16}$O. 

    The  line lists discussed above provide comprehensive coverage of ground state transitions but do not allow
    for the transitions involving different electronic states.
    LIFBASE by \citet{99LIFBASE} is an exception.
    It contains line lists for the
    $\upgamma$ (\ASigma -- \XPi), $\upbeta$ (\BPi -- \XPi), 
    $\updelta$ (\CPi -- \XPi) and $\upvarepsilon$ (\DSigma -- \XPi) NO band systems
    and provides relative cross sections via an interactive front-end program.
    However, there are some issues with the LIFBASE database, which appears to be no longer maintained, such as 
    some bands of high intensities (e.g. $\upbeta(12,0)$ and $\updelta(2,0)$)
    are missing.
    Due to the strong interactions and curve crossings between its electronically excited state,
    it is not easy to model the excited states of NO.
    Similar interactions 
    were also reported in the works discussing other
    nitrogen oxide  e.g. \ce{NO3}
    \citep{07Stanto.NO3}.
    To address this problem,
    we  \citep{jt821} proposed a method based on
    directly diagonalizing a rovibronic matrix using the variational nuclear motion program \duo \citep{jt609}. 
    Our model used a diabatic representation to resolve 
    the energy structures of \BPi--\CPi coupled states.
    A \marvel (measured active rotation vibration energy levels) analysis \citep{jt421,jt750}
    was used to produce a set of empirical energy levels of spectroscopic accuracy.
    These levels were then used by \textsc{Duo} to produce curves and couplings that give an accurate
    description of the \ASigma, \BPi and \CPi states for levels up to \SI{63000}{\per\cm} above the ground state; 
    the previous \noname\ line list already provides
    a good spectroscopic model for the \XPi state.
    This new spectroscopic model provides an excellent starting point 
    for computing a line list for
    the  $\upgamma$, $\upbeta$ and $\updelta$ band systems of NO. 

\begin{figure}
        \centering
        \includegraphics{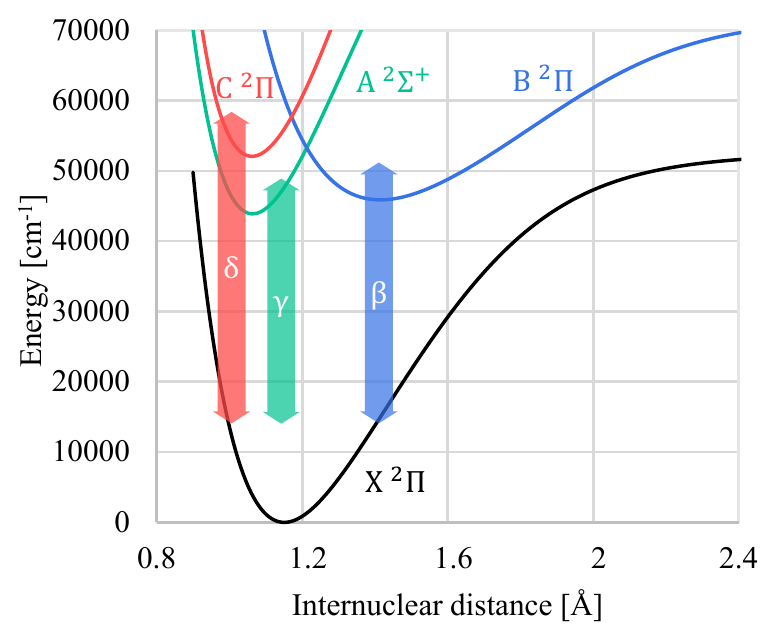}
        \caption{The band systems of 
        NO involved in this work and their names.
        For a comprehensive diagram,
        see \citet{00CaBrCa.NO}. 
        }
        \label{fig:bandSystem}
    \end{figure}

    This work aims to provide
    a line list which is both accurate and complete covering 
    the rovibrational transitions within the \XPi state
    and the  rovibronic transitions belonging to the 
    other three important band systems,
    i.e.,
    $\upgamma$, 
    $\upbeta$  and $\updelta$
    as shown in Fig.\,\ref{fig:bandSystem}. 
    Rovibronic energy levels and wavefunctions
    were taken from our published model \citep{jt821}. The TDMC  of 
    \ASigma -- \XPi was taken from the literature \citep{99LuCrxx.NO}, while the \BPi -- \XPi and     \CPi -- \XPi TDMCs, involving two strongly coupled electronic states \BPi and \CPi,  
    are constructed in the present work.
To make the line list more accurate, some calculated values are replaced with empirical ones and the associated uncertainties were provided. 
These issues are discussed in turn in the following sections

\section{Transition dipole moments}
\label{sec:TDMC}
\begin{figure*}
        \centering
        \includegraphics{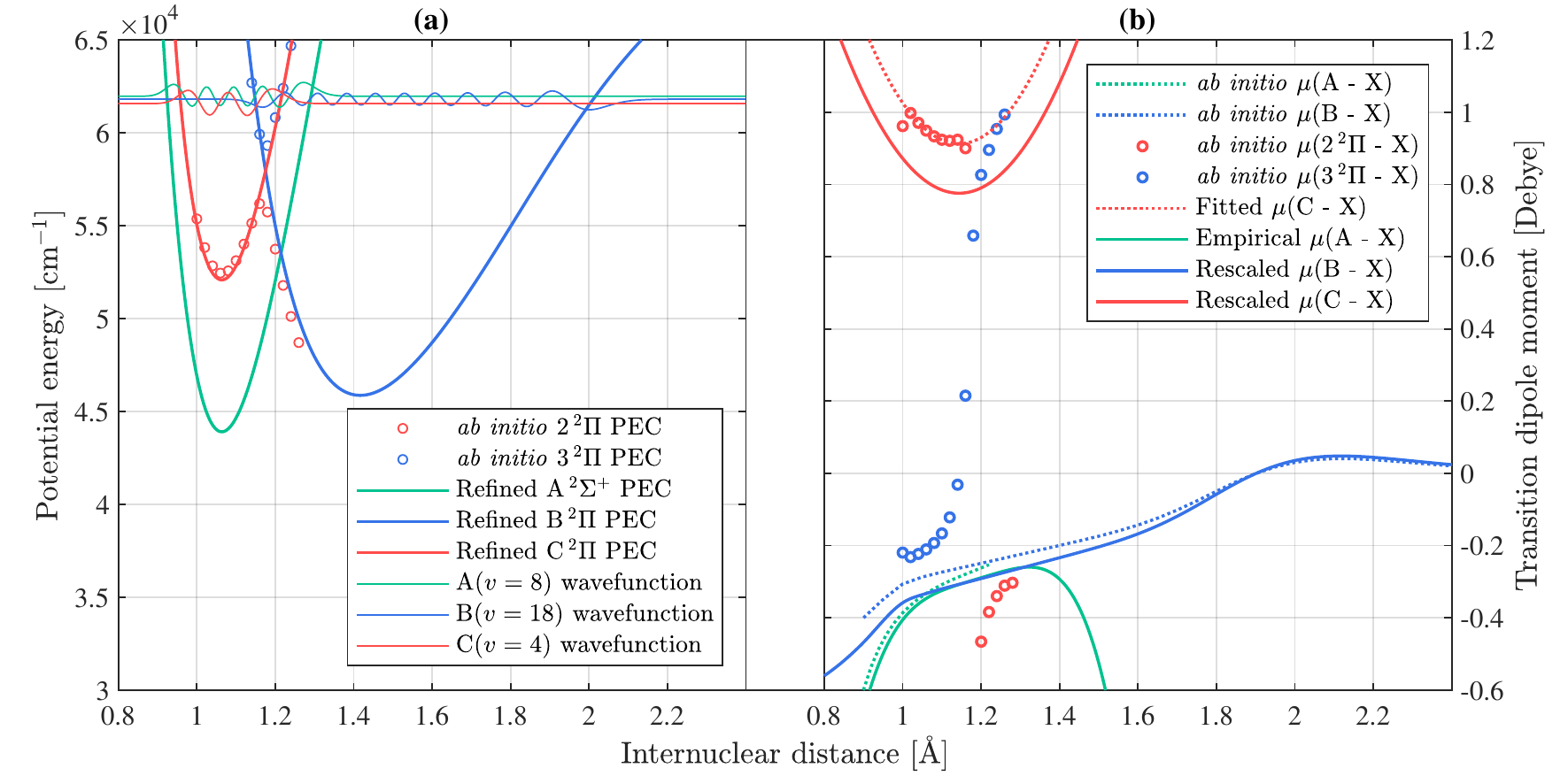}
        \caption{(a) \textit{Ab initio} and refined PECs
            as well as vibrational wavefunctions and
            (b) corresponding transition dipole moments, $\mu(\cdot)$.
            In Panel (a),
            the wavefucntions are plotted in arbitrary units.
            $2\,^2\Pi$ is the adiabatic \CPi to \BPi state
            and $3\,^2\Pi$ is the adiabatic \BPi to \CPi state.
            In Panel (b),
            `Fitted $\mu(C-X)$' is a quadratic polynomial
            which was fitted to the 
            values of red and blue circles it passes through.
            `Empirical $\mu(A-X)$' was calculated with the
            parameters determined by \citet{99LuCrxx.NO}.
            }
        \label{fig:compareDipole}
    \end{figure*}

    Our previous study \citep{jt821} explored the difficulty of preforming
    \abinitio\ calculations of NO  in some detail; 
    fundamentally the problem arises because the \ASigma and \CPi states
    are effectively Rydberg-like in character which means their curves follow that of the tightly bound \ce{NO+} ion
    \citep{98Prattx.NO}, 
    while the \BPi state is a valence state with a much flatter curve which crosses the others, see Fig.\,\ref{fig:compareDipole} (a).
    Discontinuities arise in the potential energy curves (PECs) and other curves
    due to the state (avoided) crossings and  interactions.
    The quality of transition dipole moment curves (TDMC)
    is strongly affected by the complicated behaviour of the excited state wavefunctions which were computed at the
    complete active space self-consistent field (CASSCF) and multi-reference configuration interaction (MRCI) levels using MOLPRO \citep{MOLPRO}.
    We therefore
    adopted a pragmatic approach to determining the TDMCs in which 
    the TDMCs were modified through a comparison with experimental data.

\subsection{Range of calculation}    
    The calculation setup is
    consistent with those of
    \citet{jt686} and
    our previous work \citep{jt821}:
    the internuclear distance, $R$, varied from
    \SIrange{0.6}{4}{\angstrom}, 
    which in \duo was discretised by 701 uniformly spaced grid points as part of the sinc DVR 
    (discrete variable representation) basis set.
    In the final calculations, 60, 15, 30, and 10 contracted vibrational basis functions were retained 
    for the \XPi, \ASigma, \BPi and \CPi states, respectively.

    As we do not include interactions with higher electronic states \citep{82GaDrxx.NO} in our model,  the highest vibrational levels of the \ASigma, \BPi and \CPi states were
    limited to 8, 18, and 4, respectively.
    The vibrational wavefunctions of these levels 
    are shown in Fig.\,\ref{fig:compareDipole} (a),
    where they are vertically shifted to their vibrational energies.
    The global energy limit was chose to be \SI{65000}{\per\cm}.

\subsection{\ASigma state}

For the \ASigma -- \XPi transitions,
we used the empirical TDMC constructed by \citet{99LuCrxx.NO} as a fourth-order polynomial (see  the green solid curve shown in Fig.\,\ref{fig:compareDipole} (b).
    We chose not to use an \abinitio TDMC of \ASigma -- \XPi
    although the one depicted by the  green dash curve 
    in Fig.\,\ref{fig:compareDipole} (b) looks very similar to the empirical one.
    We found that use of different active spaces in MOLPRO
    gave TDMCs that were very different in both shape and amplitude; this behaviour  is 
    discussed by \citet{94ShBaLa.NO}.
    We thus had to compare these curves with the empirical TDMC and
    select a similar one; this
     procedure is neither `\abinitio' nor `empirical'.
    The empirical TDMC function of \citet{99LuCrxx.NO} was based on 
    the lower vibrational levels of the \ASigma state. As a result,
    their polynomial diverges  at distances $R>\SI{1.3}{\angstrom}$.
    We chose to use this TDMC unaltered as the vibrational wavefunctions 
    decay rapidly to zero in this region so our line list 
    is insensitive to the behaviour of the TDMC at these values of $R$.

\begin{figure*}
        \centering
        \includegraphics{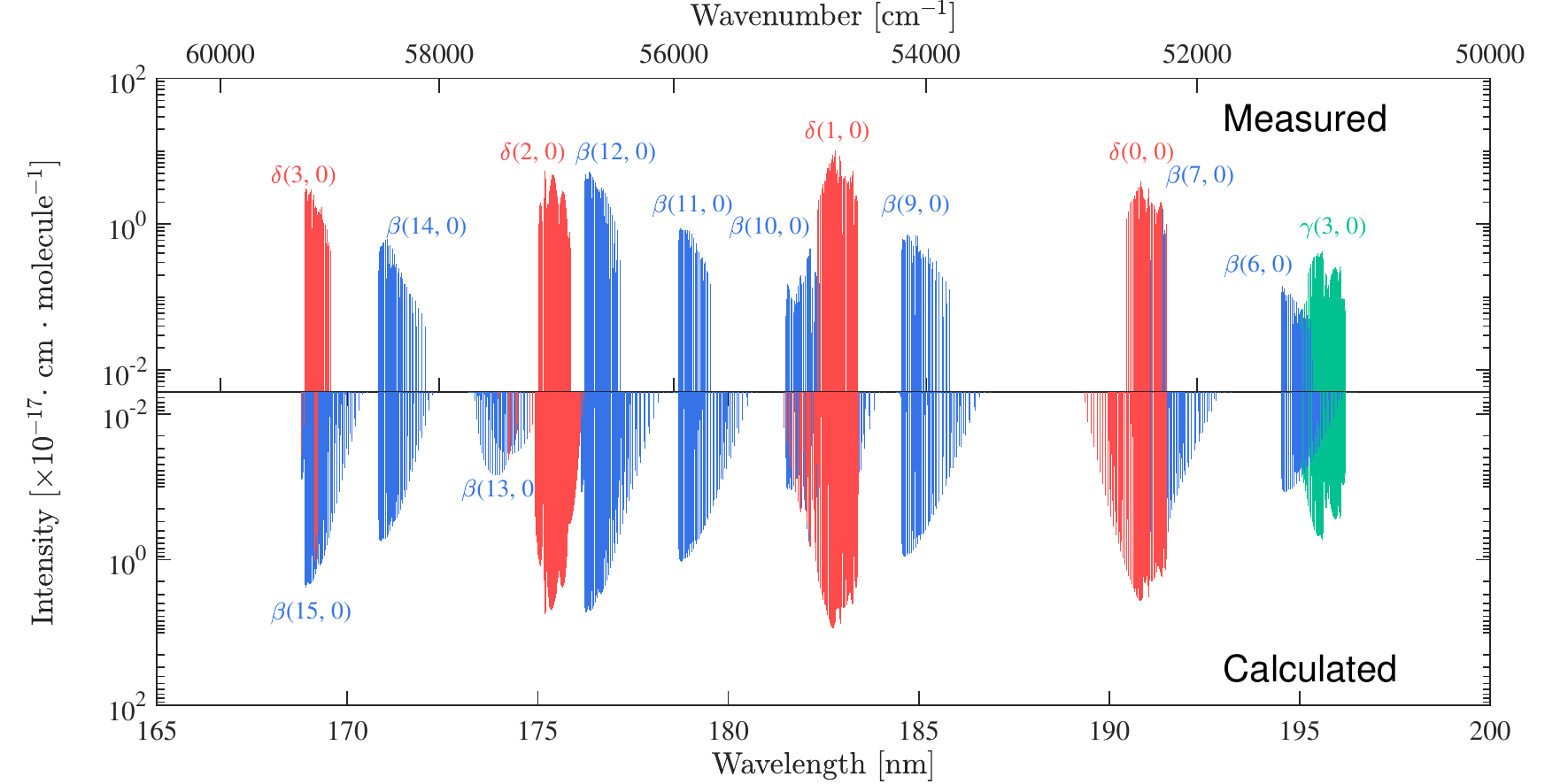}
        \caption{Calculated absorption intensities at 295 K  
        compared with the values given by \citet{06YoThMu.NO}.
       As the fine-structure doublets 
        for more than half the transitions 
        were not resolved in the experiment, 
    all doublets are removed by averaging  the positions of the two lines
        and adding their intensities,
        for both measured and calculated
        transitions.
        }
        \label{fig:YoIntensity}
    \end{figure*}

\citet{09SePaHu.NO} updated the TDMC polynomial coefficients of \citet{99LuCrxx.NO} using   radiative lifetimes of NO \ASigma($v=0, 1, 2$) which they measured using time-resolved laser-induced fluorescence. Their transition dipole moment is larger in magnitude than that of Luque and Crosley. We use the TDMC of Luque and Crosley as a balance between measured radiative lifetimes \citep{99LuCrxx.NO, 09SePaHu.NO} and the intensities (\lq integrated cross section\rq) of the  $\upgamma(3,0)$ measured by 
\citet{06YoThMu.NO}; see  Fig.\,\ref{fig:YoIntensity}.

\subsection{\BPi-\CPi coupled states}

\begin{figure}
    \centering
    \includegraphics{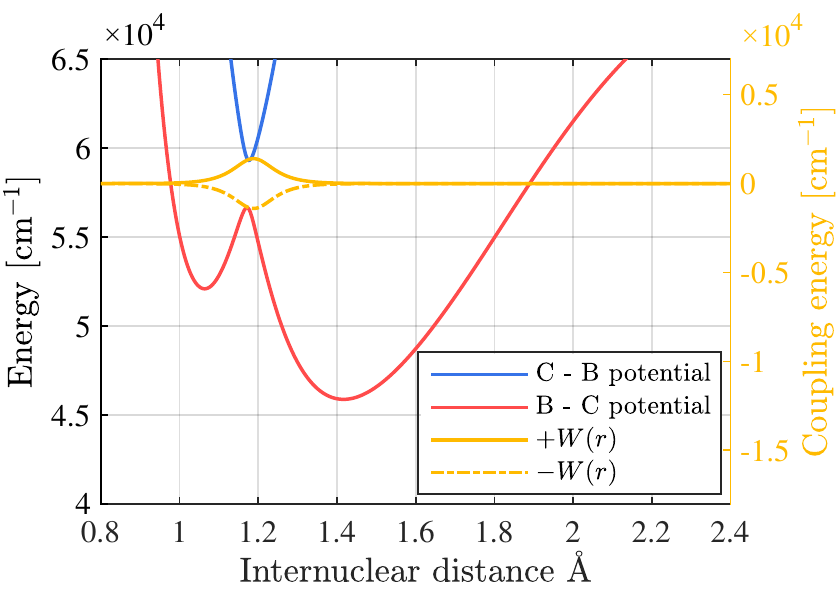}
    \caption{Eigenvalue curves of Eq.\,\eqref{eq:wr}}
    \label{fig:aidabaticPEC}
\end{figure}

The adiabatic PECs \BPi and  \CPi have the same symmetry and 
therefore form  an avoided crossing as shown by  our  \abinitio\ PECs computed using MOLPRO,
see circles in  Fig.~\ref{fig:compareDipole}(a).  
In order to avoid discontinuities in the various curves, 
including the TDMCs, 
here we  follow our diabatic model \citep{jt821} with the PECs shown in Fig.\,\ref{fig:compareDipole} (a).
Off-diagonal matrix elements were introduced to 
represent the electronic state interaction as follows:
\begin{equation}
    \label{eq:wr}
    \begin{pmatrix}
        V_\mathrm{B}(r) & W(r) \\
        W(r) & V_\mathrm{C}(r) 
    \end{pmatrix} \, ,
\end{equation}
where $V_\mathrm{B}(r)$ and $V_\mathrm{C}(r)$
are two diabatic potentials and $W(r)$ is
a bell-shaped coupling curve. The effect of this representation is illustrated in in Fig.\,\ref{fig:aidabaticPEC}, 
although the above matrix elements are introduced into the  rovibronic Hamiltonian matrix rather than directly diagonalizing Eq.~\eqref{eq:wr} to generate adiabatic potentials  shown in this figure. 

Due to the  (adiabatic) avoided crossing between \BPi and \CPi, 
the  adiabatic TDMCs  \BPi -- \XPi and \CPi -- \XPi  exhibit erratic  behaviour 
in the interaction region with their  values  rising  and dropping  sharply near \SI{1.18}{\angstrom}, 
as shown by the blue and red circles  in Fig.\,\ref{fig:compareDipole} (b). 
The interaction center (about \SI{1.18}{\angstrom}) is close to the  
equilibrium bond length (about \SI{1.15}{\angstrom}) of the \XPi state. 
Thus, even a slight change in the TDMC in this region along 
internuclear distance axis can dramatically change the calculated transition intensities.

In the  diabatic model, the  \BPi -- \XPi and     \CPi -- \XPi TDMC are  smooth  curves (see Fig.~\ref{fig:compareDipole} (b)) which do not show erratic behaviour in the interaction region and therefore are no longer the most sensitive factor 
in intensity calculations. The coupling between states in this model is  controlled  by and relies on the quality of the  rovibronic \BPi and  \CPi wavefunctions, which can be accurately determined by our technique of fitting theoretical curves using experimental energies. 
    
The original \abinitio PECs for \XPi and \BPi and the corresponding TDMC of \BPi -- \XPi were computed using a high level of theory,  CASSCF\&MRCI+Q/cc-pVQZ with the  $[(6, 2, 2, 0) - (2, 0, 0, 0)]$
active space, where $[(n_1, n_2, n_3, n_4) - (n'_1, n'_2, n'_3, n'_4)]$ indicates the occupied and closed orbitals in the irreducible representations $a_1$, $b_1$, $b_2$ and  $a_2$ of the C$_{2v}$ point group (shown as red and blue circles in Fig.\,\ref{fig:compareDipole}). These adiabatic data  were then diabatized 
to produce curves  shown in Fig.\,\ref{fig:compareDipole} as  dotted curves. In order to improve the quality of the intensity calculations, the diabtaic TDMCs were then further empirically adjusted as follows.  
The \BPi -- \XPi TDMC  was scaled using a combination of the measured lifetimes of \citet{95LuCrxx.NO} and integrated cross sections of \citet{06YoThMu.NO}. 
The  scaling factor of \num{1.17} was chosen as a  balance between these two experiments.
The scaled \BPi -- \XPi TDMC  is shown as a blue solid line  in Fig\,\ref{fig:compareDipole}. 
Although the \BPi -- \XPi TDMC diverges from its original trend for the internuclear distances shorter than \SI{1}{\angstrom}, this does not affect our calculation 
as the corresponding \BPi  vibrational wavefunctions  nearly vanish there. 
    
The TDMC of \CPi -- \XPi required special care; 
we first fitted a quadratic polynomial to the adiabatic
\abinitio values,  
shown as the red dash curve in Fig.\,\ref{fig:compareDipole}, and then scaled it based on the absorption intensities measured by \citet{06YoThMu.NO}.
Since the TDMC of \CPi -- \XPi influences the intensities of higher $\upbeta$ bands (i.e. $v_{\rm B} \ge 7$), it was important to obtain a  global agreement for all intensities, including the $\upbeta$  system. This is illustrated in  Fig.\,\ref{fig:YoIntensity}, where the calculated spectrum of NO in the region of 
\SI{200}{\nm} is compared to the experimental intensities by \citet{06YoThMu.NO}. 
    
All curves comprising our spectroscopic mode, including the TDMCs, are provided in the supplementary material to this paper as a \duo input file.

\section{Line list calculation}

A rovibronic line list \XABC for NO was constructed using the \duo program and the spectroscopic model described above. For the detailed description of the \duo calculation see our
previous work \citep{jt821}.

    The \XABC line list consists of \num{4596666}  transitions 
    between \num{30811} states of the four low-lying 
    electronic states \XPi, \ASigma, \BPi and \CPi, (\num{21668}, 
    \num{1209}, \num{6873} and \num{1041}, respectively), 
    covering $J \leq 184.5$ with rovibronic wavenumber cutoffs   
    of \SI{53000}{\per\cm} (\XPi, same as in \citet{jt686}) and  \SI{63000}{\per\cm} (all other states). 
    These energy cutoffs are smaller than the basis set limit  
    to avoid any truncation problems near \SI{63000}{\per\cm}. 
    In line with the   ExoMol data structure \citep{jt548}, the line list is represented by two files, a \texttt{.states} (states) file
    and a \texttt{.trans} (transitions) file.   
    Table\,\ref{tab:state} gives an extract of the \XABC \texttt{.states} file. 
The \texttt{.trans} (transitions) file contains the  Einstein-$A$ coefficients 
calculated with the \duo spectroscopic model of this work
and constitutes our new \XABC line list.     
Table\,\ref{tab:transition} gives an extract from the \texttt{.trans}  file.

\begin{figure}
        \centering
        \includegraphics{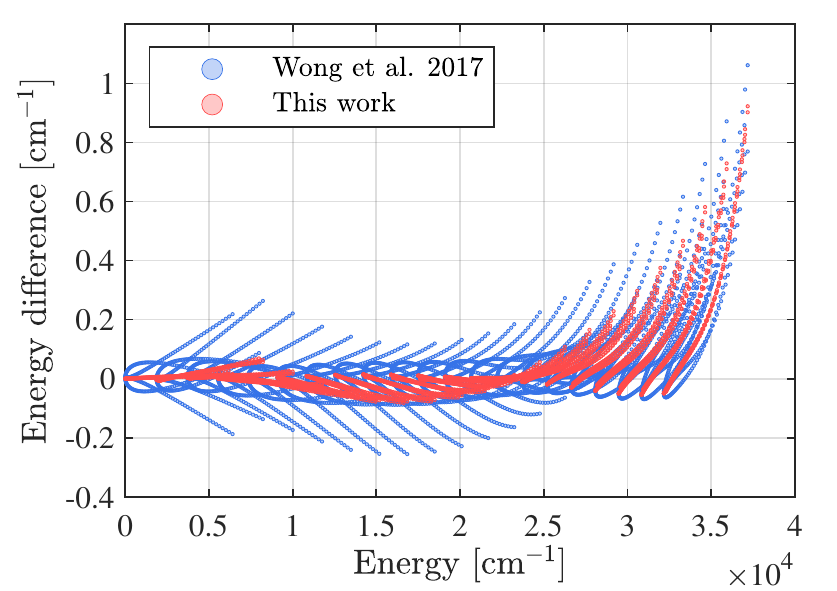}
        \caption{
        Energy differences between
        the results of \duo and SPFIT/SPCAT 
        for $J \leq 60.5$, $v\leq 20$ states of \XPi.
        }
        \label{fig:spfitdiff}
    \end{figure}

    The current spectroscopic model uses improved \Ldoubling parameters for the \XPi state compared to \citet{jt686}. As a consequence,  the new model shows better agreement with the effective Hamiltonian  SPFIT/SPCAT energies of NO also presented in \citet{jt686}, see Fig.\,\ref{fig:spfitdiff}. 
    Due to the change in the model and consequently the wavefunctions,  
    the Einstein-$A$ coefficients between the states of \XPi 
    as well as the corresponding lifetimes  have also changed. 
    The energies of the  lower rovibronic states ($v \leq 29$ and $J \leq 99.5$) of \XPi 
    were replaced with the \noname ones which were
    calculated by \citet{jt686} using the programs SPFIT and SPCAT \citep{SPFIT},
    based on the work of \citet{15MuKoTa.NO}.
    These states are  labeled with \texttt{EH} to indicate that  
    they were calculated from effective Hamiltonian, which
    replaces the label \texttt{e} (i.e. empirical)
    used in \noname for these \texttt{EH} levels.
    The energies of the other states of \XPi
    were shifted from the results of \duo, 
    using the same strategy as \citet{jt686}, 
    to avoid energy jumps
    above $v=29$ or $J = 99.5$.
    These shifted states were labeled with \texttt{Sh}
    while they were labeled with \texttt{c} (i.e. calculated) in \noname.

    We also replaced the \duo energies of the \ASigma, \BPi and \CPi states 
    with \marvel energies where available,
    these states are labeled \texttt{Ma} in the \texttt{.states} file.  
    The jumps between \marvel and \duo energies 
    in the excited electronic states are negligible. 
    Therefore,  we did not shift the other calculated energies 
    of the \ASigma, \BPi or \CPi states
    and labeled these states with \texttt{Ca}.

\begin{figure}
        \centering
        \includegraphics{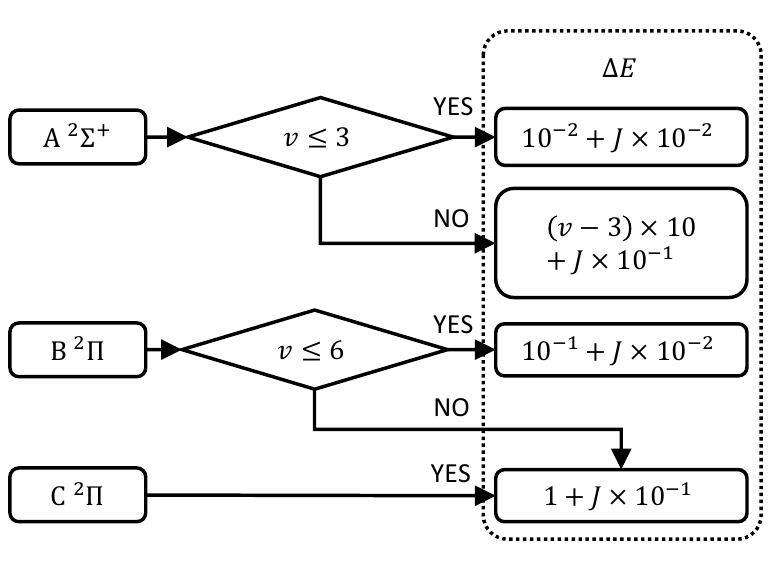}
        \caption{The uncertainties assigned to
        the calculated energies of \ASigma,
        \BPi and \CPi in \si{\per\cm}.}
        \label{fig:uncertainty_abc}
    \end{figure}

\begin{figure}
        \centering
        \includegraphics{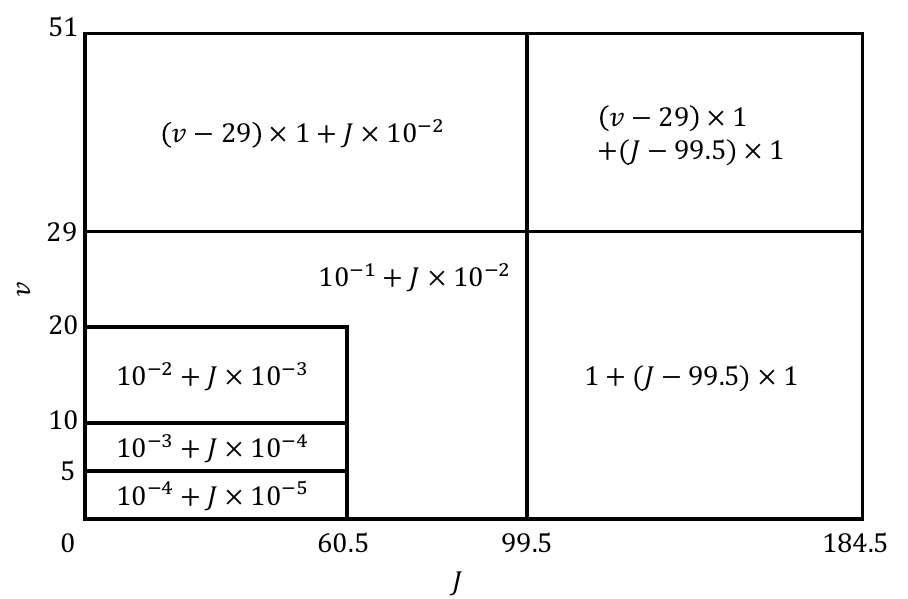}
        \caption{Uncertainties assigned to the 
        energy levels of \XPi state in \si{\per\cm}.
        The values are consistent with
        the recent  HITEMP Uncertainty Codes  given for  
        NO  \citep{jt763}.}
        \label{fig:uncertainty_x}
    \end{figure}

    For this line list we also introduced an extra column 
    containing the energy uncertainties 
    (i.e. Column 5, $\Delta E$) for each state. 
    The  uncertainties of  \ASigma, \BPi and \CPi  were
    taken from MARVEL analysis where available.
    Otherwise, they  were constructed  
    according to corresponding vibrational and rotational quantum numbers 
    using the algorithm  shown in Fig.\,\ref{fig:uncertainty_abc}. 
    The estimation of  uncertainties of the \XPi state is a bit more complicated but is  based on the same idea, see Fig.\,\ref{fig:uncertainty_x}.

\begin{landscape}    
    \begin{table*}
        \caption{Extract from NO \XABC \texttt{.states} file. }
        \label{tab:state}
        \tt
        \begin{tabular}{rcccrcrcclrccccr}
        \hline
        $i$ & $E$ & $g_i$ & $J$ &\multicolumn{1}{c}{$\Delta E$} & $\tau$ &\multicolumn{1}{c}{$g_J$} & $+/-$ & $e/f$    & \textrm{state} & $v$ & $\varLambda$ & $\varSigma$ & $\varOmega$ & \textrm{label} &\multicolumn{1}{c}{$E_\duo$} \\
        \hline
        61    & 51869.286798  & 6     & 0.5   & 20.000000  & 4.8551E-02 & -0.000767  & +     & e     & X2Pi  & 49    & 1     & -0.5  & 0.5   & Sh    & 51871.759731  \\
        62    & 51970.104351  & 6     & 0.5   & 21.000000  & 4.3477E-02 & -0.000767  & +     & e     & X2Pi  & 50    & 1     & -0.5  & 0.5   & Sh    & 51972.577284  \\
        63    & 52081.384882  & 6     & 0.5   & 22.000000  & 3.9557E-02 & -0.000767  & +     & e     & X2Pi  & 51    & 1     & -0.5  & 0.5   & Sh    & 52083.857815  \\
        64    & 52345.934940  & 6     & 0.5   & 1.050000  & 5.1915E-07 & -0.000767  & +     & e     & B2Pi  & 7     & 1     & -0.5  & 0.5   & Ca    & 52345.934940  \\
        65    & 52372.741033  & 6     & 0.5   & 0.007071  & 4.4075E-08 & -0.000767  & +     & e     & C2Pi  & 0     & 1     & -0.5  & 0.5   & Ma    & 52372.500922  \\
        66    & 53273.412094  & 6     & 0.5   & 0.141421  & 8.2023E-07 & -0.000767  & +     & e     & B2Pi  & 8     & 1     & -0.5  & 0.5   & Ma    & 53273.523369  \\
        67    & 53370.608307  & 6     & 0.5   & 10.050000  & 1.8421E-07 & 2.002313  & +     & e     & A2Sigma+ & 4     & 0     & 0.5   & 0.5   & Ca    & 53370.608307  \\
        68    & 54183.455941  & 6     & 0.5   & 0.020000  & 5.8924E-07 & -0.000767  & +     & e     & B2Pi  & 9     & 1     & -0.5  & 0.5   & Ma    & 54183.325295  \\
        69    & 54690.017247  & 6     & 0.5   & 0.030000  & 4.0148E-08 & -0.000767  & +     & e     & C2Pi  & 1     & 1     & -0.5  & 0.5   & Ma    & 54689.759899  \\
        70    & 55090.440941  & 6     & 0.5   & 0.030000  & 5.1073E-07 & -0.000767  & +     & e     & B2Pi  & 10    & 1     & -0.5  & 0.5   & Ma    & 55090.115548  \\
        \hline
        $i$ & \multicolumn{10}{l}{\textrm{Counting number}} \\
        $E$& \multicolumn{10}{l}{\textrm{State energy in} \si{\per\cm}}\\
        $g_i$& \multicolumn{10}{l}{\textrm{Total state degeneracy}}\\
        $J$& \multicolumn{10}{l}{\textrm{Total angular momentum}} \\
        $\Delta E$& \multicolumn{10}{l}{\textrm{Energy uncertainty in} \si{\per\cm}}\\
        $\tau$& \multicolumn{10}{l}{\textrm{Lifetime in} \si{\second}}\\
        $g_J$& \multicolumn{10}{l}{\textrm{Lande $g$-factor}}\\
        $+/-$& \multicolumn{10}{l}{\textrm{Total parity}}\\
        $e/f$& \multicolumn{10}{l}{\textrm{Rotationless parity}} \\
        \textrm{state}& \multicolumn{10}{l}{\textrm{Electronic state}}\\
        $v$& \multicolumn{10}{l}{\textrm{Vibrational quantum number}}\\
        $\varLambda$& \multicolumn{10}{l}{\textrm{Projection of electronic angular momentum}}\\
        $\varSigma$& \multicolumn{10}{l}{\textrm{Projection of the electronic spin}}\\
        $\varOmega$& \multicolumn{10}{l}{\textrm{Projection of the total angular momentum}}\\
        \textrm{label}& \multicolumn{10}{l}{
        Sh \textrm{for shifted,}
        Ca \textrm{for calculated,} 
        EH \textrm{for effective Hamiltonian,}
        Ma \textrm{for \marvel}}\\
        $E_\duo$& \multicolumn{10}{l}{\textrm{State energy in \si{\per\cm} calculated  with \duo}}\\
        \end{tabular}
    \end{table*}
\end{landscape}

\begin{figure}
        \centering
        \includegraphics{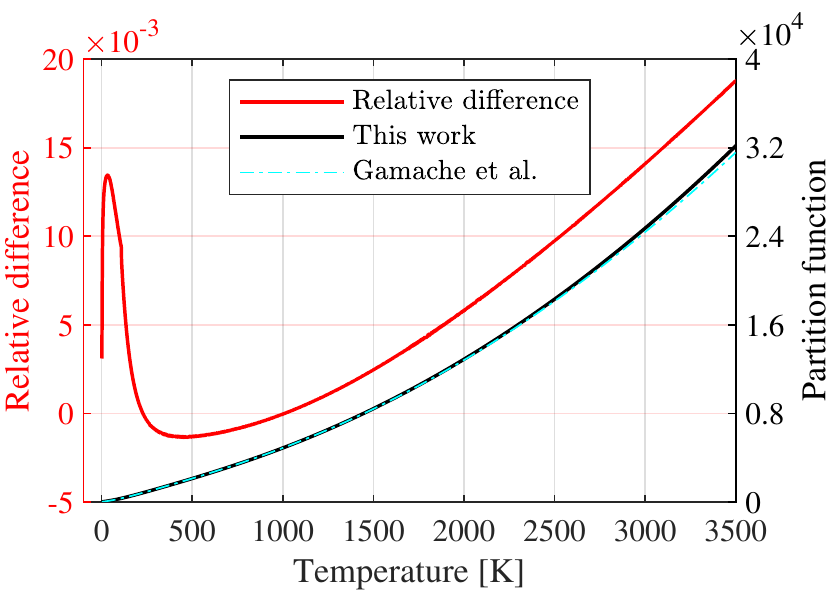}
        \caption{Partition function calculated 
            using the XABC state energies
            in comparison with the TIPS values
            of \citet{jt692}.
            The red curve illustrates the 
            relative difference between them.}
        \label{fig:partfunc}
    \end{figure}

The partition function was computed  using the standard summation 
over energies using  levels  from our final line list.
Figure\,\ref{fig:partfunc} compares the Total Internal Partition Sums (TIPS) partition functions, $Q(T)$,  
of \citet{jt692} and this work. 
As can be seen from  the red curve,  
i.e., $(Q_\XABC - Q_\mathrm{Gamache}) / Q_\mathrm{Gamache}$,
the partition function difference is very small because the thermodynamic properties are generally not very sensitive to small variations of higher-lying energies as in  the \XPi state, while the  \ASigma, \BPi and \CPi energies (not considered in TIPS) are too high to make 
obvious difference for the temperature below \SI{3500}{K} considered here.

     \begin{table}
        \caption{Extract from the NO \XABC \texttt{.trans} file.}
        \label{tab:transition}
        \tt
        \begin{tabular}{rrcc}
        \hline
        $f$     & $i$     & $A_{fi}$  & $\nu_{fi}$ \\
        \hline
        117   & 1     & 3.1174E+05 & 44203.001767  \\
        129   & 1     & 6.4207E+05 & 48854.091482  \\
        162   & 1     & 2.0740E+05 & 58538.988402  \\
        151   & 1     & 1.0243E+05 & 53273.417513  \\
        156   & 1     & 2.5855E+04 & 55586.090674  \\
        150   & 1     & 1.4370E+06 & 52372.686268  \\
        149   & 1     & 2.3080E+05 & 52345.937733  \\
        134   & 1     & 1.4276E+04 & 50452.598447  \\
        122   & 1     & 1.3786E+02 & 46503.348087  \\
        168   & 1     & 5.2902E+05 & 61721.081024  \\
        \hline
        $f$ &\multicolumn{3}{l}{\textrm{Counting number of the upper state}}\\
        $i$&\multicolumn{3}{l}{\textrm{Counting number of the lower state}}\\
        $A_{fi}$&\multicolumn{3}{l}{\textrm{Einstein-$A$ coefficient in} \si{\per\second}}\\
        $\nu$&\multicolumn{3}{l}{\textrm{Transition wavenumber in cm$^{-1}$.}}\\
        \end{tabular}
    \end{table}
    
\begin{figure*}
        \centering
        \includegraphics{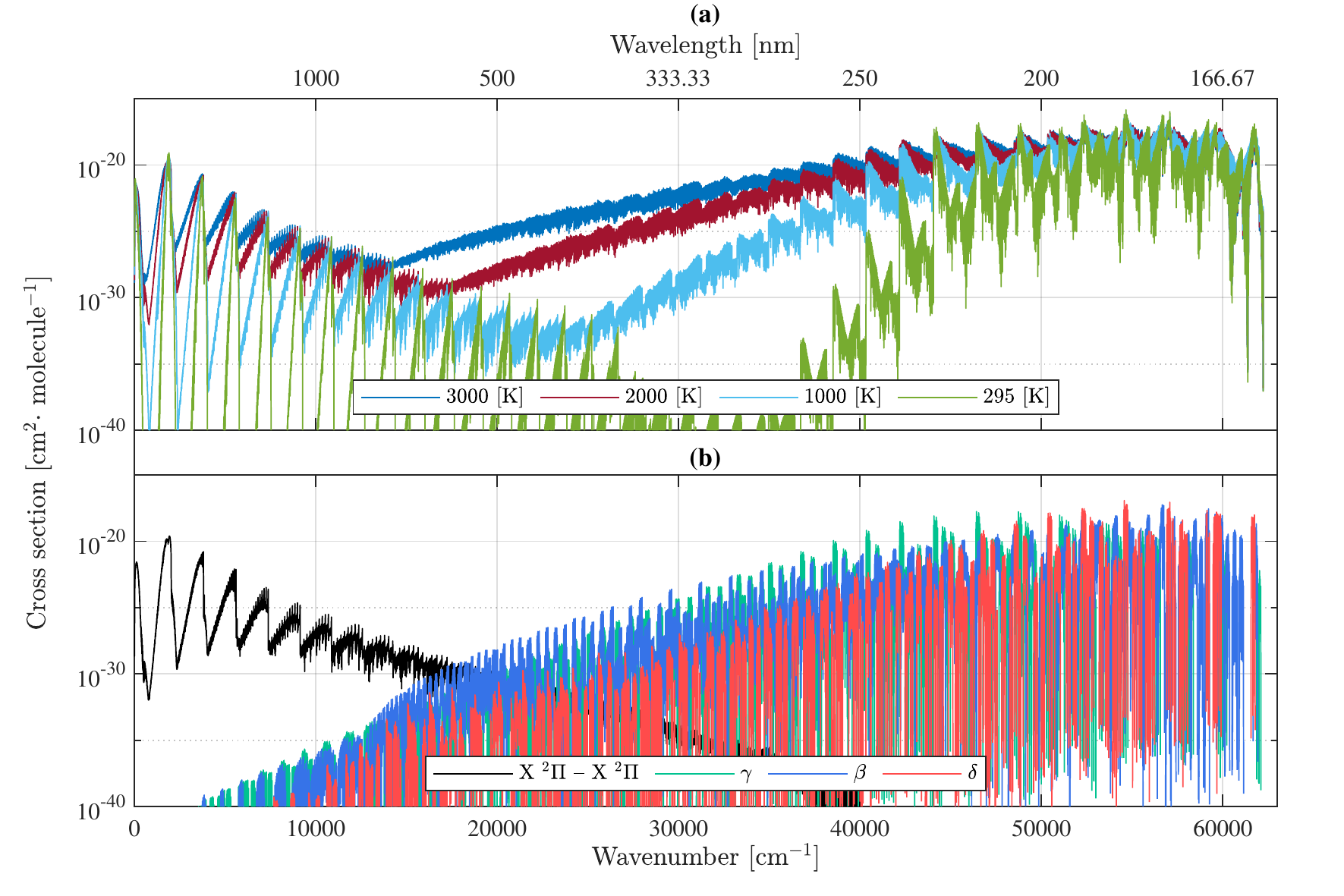}
        \caption{
        \NO cross sections
        below \SI{63000}{\per\cm}
        calculated using the \XABC line list and a Gaussian lineshape
        function with a HWHM of \SI{1}{\per\cm}:  
        (a) Calculated cross sections of NO at different temperatures; 
        (b) \XPi -- \XPi, $\upgamma$, $\upbeta$ and $\updelta$ cross sections at \SI{2000}{K}.}
        \label{fig:crossAll}
    \end{figure*}

With the assumption of local thermal equilibrium, we calculated  absorption spectra of NO using the new line list \XABC. An overview of  the \XABC absorption spectrum fo different temperatures below \SI{63000}{\per\cm} or longward \SI{1600}{\angstrom}
is shown in Fig.\,\ref{fig:crossAll}.
    
 \section{Comparisons}  
 
    Unless otherwise indicated, 
    the following calculations were executed with \exocross \citep{jt708},
    which is a program 
    for generating lifetimes, spectra, partition function etc.,
    from molecular line lists.
\subsection{Lifetimes}    
    
\begin{figure*}
        \centering
        \includegraphics{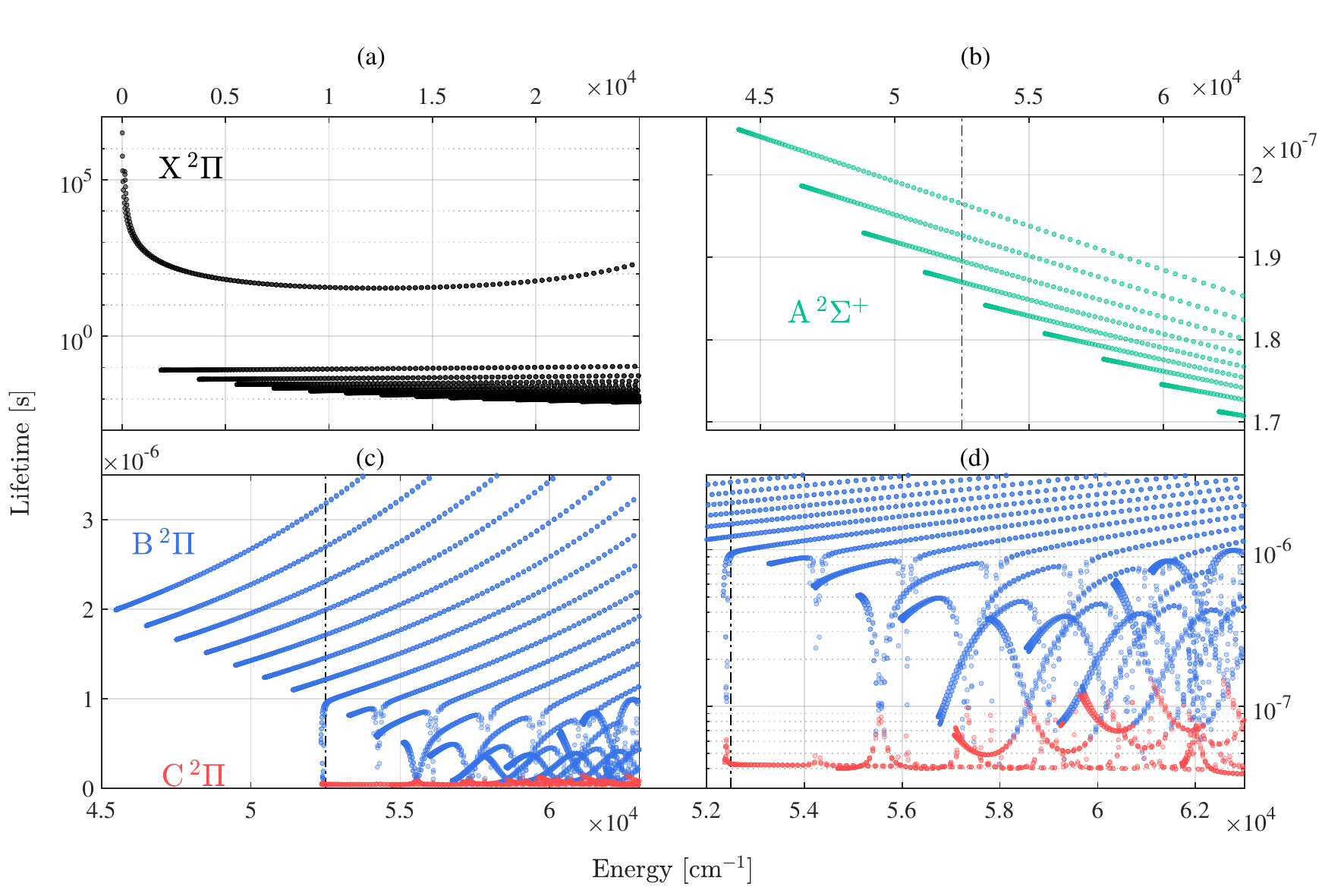}
        \caption{Calculated lifetimes
            of (a) \XPi, (b) \ASigma, 
            and (c)\&(d) \BPi-\CPi coupled states.
            The lifetimes of the two lowest states 
            (\XPi, $v=0$, $J=1/2$, $\varOmega = 1/2$,
            $e/f$)
            are respectively infinity and 2.3$\times 10^{14}$~s
            and are omitted from Panel (a).            
            The vertical dot-dash lines in
            Panels (b), (c) and (d) indicate
            the first dissociation limit of NO.
            Panel (d) is a blow up of Panel (c).
            }
        \label{fig:NOlife}
    \end{figure*}

Lifetimes for individual states of \XPi, \ASigma, \BPi and \CPi are plotted against the energies in Fig.\,\ref{fig:NOlife}. The vibronic lifetimes of \ASigma($v=0$ to $3$) and  \BPi ($v=0$ to $6$) are compared with experimental values (where available) in  Tables \ref{tab:Alife} and \ref{tab:Blife}, respectively. The calculated lifetimes of \ASigma state  agree well with those measured by \citet{99LuCrxx.NO}.  As we used their TDMC, the agreement means that \duo gave similar  vibrational wavefunctions   as the RKR (Rydberg– Klein–Rees) ones they used. Our computed lifetimes for the \BPi state are larger than those of previous works.

The funnel-like shapes of the dependence of lifetimes on the energy shown in Fig.\,\ref{fig:NOlife} (b)  are caused by the interactions between the \BPi and \CPi vibronic levels.  The lifetimes decrease to much smaller values  when rotational quantum numbers are trapped in these funnels.  Apart from the electronic state interaction, the observed lifetimes are further shortened by predissociation. 
The dot-dash lines in Fig.\,\ref{fig:NOlife}  illustrate the first dissociation limit of NO.  As the current version of \duo  does not allow for predissociation,   the calculated lifetimes to the right of the dot-dash lines are expected to be larger than the observed ones. For example,  the calculated lifetimes of 
\CPi are    of order   \SI{10}{ns}   whereas the measured ones can be    as short as several nanoseconds  \citep{89HaBoxx.NO}.
    
    \begin{table}
        \caption{Vibronic radiative lifetimes for the \ASigma state.}
        \label{tab:Alife}
        \begin{tabular}{ccccc}
        \hline
        \multirow{2}[4]{*}{$v$} & \multicolumn{2}{c}{Measured [\si{\ns}]} & \multicolumn{2}{c}{Calculated [\si{\ns}]} \\
              & Ref.$^a$ & Ref.$^b$ & Ref.$^c$ & This work \\
        \hline
        0     & $205\pm \ \ 7$ & $192.6\pm0.2$ & 206   & 205.5 \\
        1     & $200\pm \ \ 7$ & $186.2\pm0.4$ & 199   & 198.6 \\
        2     & $192\pm \ \ 7$ & $179.4\pm0.7$ & 193   & 192.9 \\
        3     & $184\pm \ \ 7$ &       & 188   & 188.1 \\
        4     & $157\pm10$ &       & 184   & 184.1 \\
        5     & $136\pm10$ &       & 180   & 180.8 \\
        \hline
        \multicolumn{5}{l}{$^a$ \citet{00LuCrxx.NO}}\\
        \multicolumn{5}{l}{$^b$ \citet{09SePaHu.NO}}\\
        \multicolumn{5}{l}{$^c$ \citet{99LuCrxx.NO}}\\
        \end{tabular}
    \end{table}
    
    \begin{table}
        \caption{Vibronic radiative lifetimes of \BPi state}
        \label{tab:Blife}
        \begin{tabular}{cccc}
        \hline
        & \multicolumn{1}{c}{Measured [\si{\us}]} & \multicolumn{2}{c}{Calculated [\si{\us}]} \\
        {$v$} & {Ref.$^a$} & {Ref.$^b$} & {This work} \\
        \hline
        0     & 2.00    & 2.00     & 2.00 \\
        1     & 1.82  & 1.77  & 1.84 \\
        2     & 1.52  & 1.56  & 1.68 \\
        3     & 1.46  & 1.39  & 1.53 \\
        4     & 1.19  & 1.24  & 1.38 \\
        5     & 1.07  & 1.11  & 1.24 \\
        6     & 0.85  & 0.99  & 1.11 \\
        \hline
        \multicolumn{4}{l}{$^a$ \citet{90GaSlxx.NO}}\\
        \multicolumn{4}{l}{$^b$ \citet{95LuCrxx.NO}}\\
        \end{tabular}
    \end{table}

\subsection{Absorption spectra}
    
\begin{figure*}
        \centering
        \includegraphics{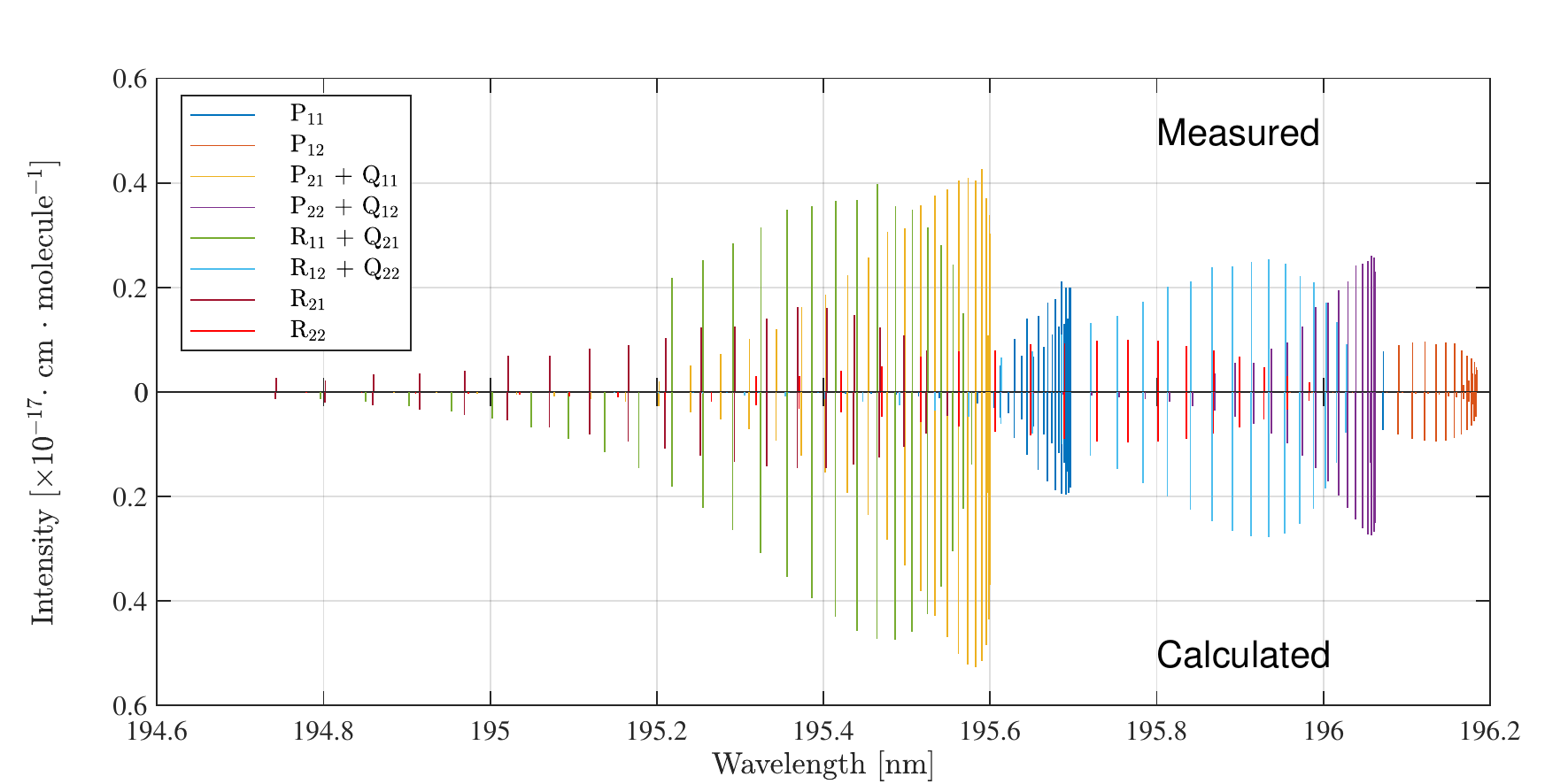}
        \caption{Calculated absorption intensities of the NO $\gamma(3, 0)$ band at 295 K
            compared with 
            the those published in the work of \citet{06YoThMu.NO}.
        As no  spin-rotational fine structure was
            observed in the experiment, 
            the wavelengths of 
            the calculated doublets are averaged and their intensities summed, to also give 
            blended lines.
            }
        \label{fig:gamma30}
    \end{figure*}

Figure\,\ref{fig:gamma30} compares the  experimental absorption intensities of $\upgamma(3,0)$ measured by \citet{06YoThMu.NO} and theoretical intensities calculated with \duo. With the TDMC of \citet{99LuCrxx.NO}, our calculations gives higher intensities than the observed ones  for the transitions of {$\mathrm{R}_{11}+\mathrm{Q}_{21}$} and  {$\mathrm{P}_{21}+\mathrm{Q}_{11}$} branches.  Thus, if we had used the TDMC of \citet{09SePaHu.NO},  \duo would further amplify the intensities of all branches, worsening agreement with observation.

\begin{figure*}
        \centering
        \includegraphics{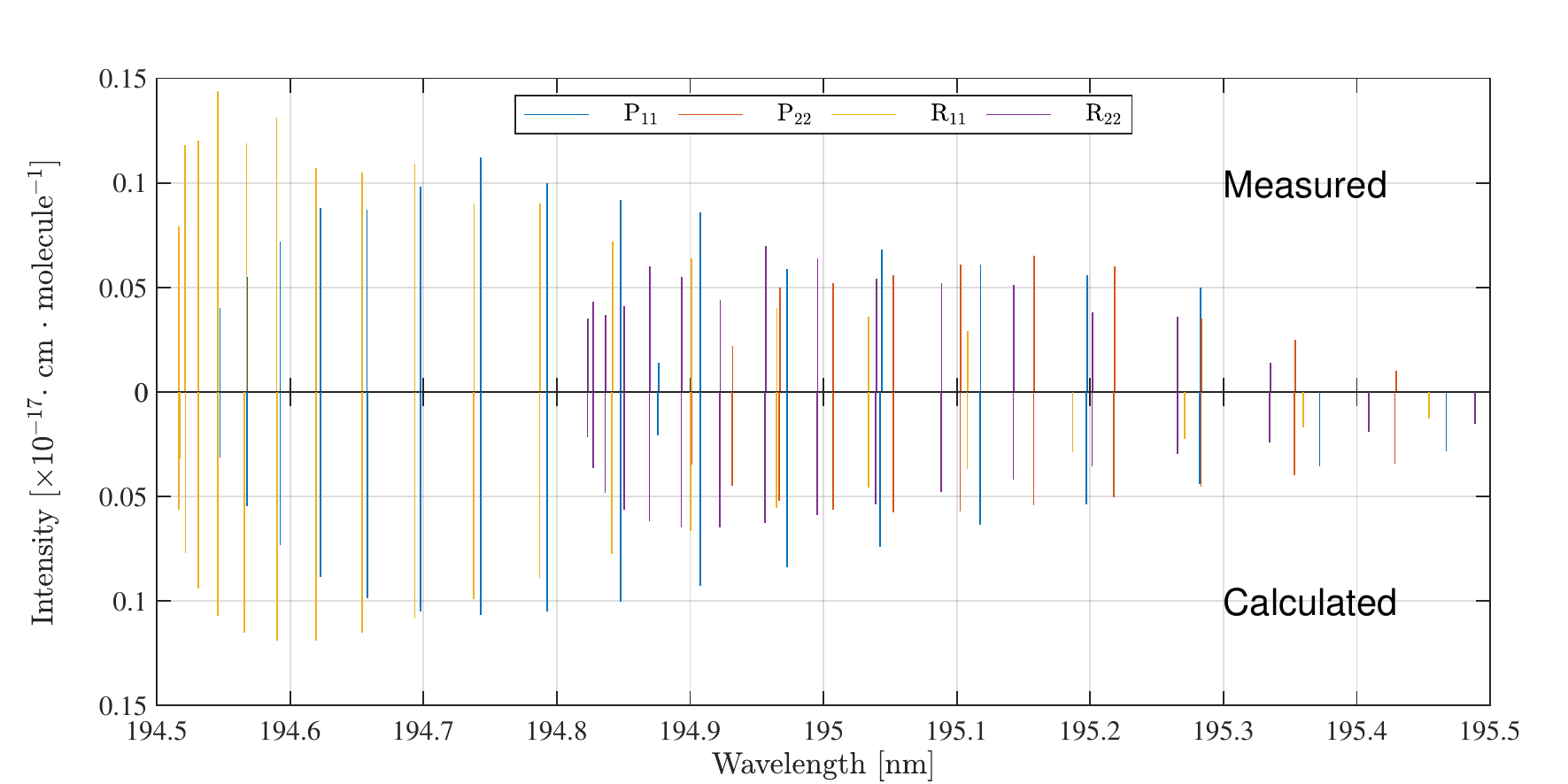}
        \caption{
            Calculated absorption intensities for the $\upbeta(6, 0)$ band at 295 K 
            in comparison with 
            the values given by \citet{06YoThMu.NO}.
            The line intensities of this band are weak
            and the experiment only resolved the
            $\varLambda$-doublets of high $J$ lines in P$_{11}$ and R$_{11}$ branches. 
            To achieve higher signal-noise ratio,
            we averaged the wavelengths of the $e$ and $f$ doublets
            and added up their intensities to create blended transitions for all branches.}
        \label{fig:beta60}
    \end{figure*}

\begin{figure*}
        \centering
        \includegraphics{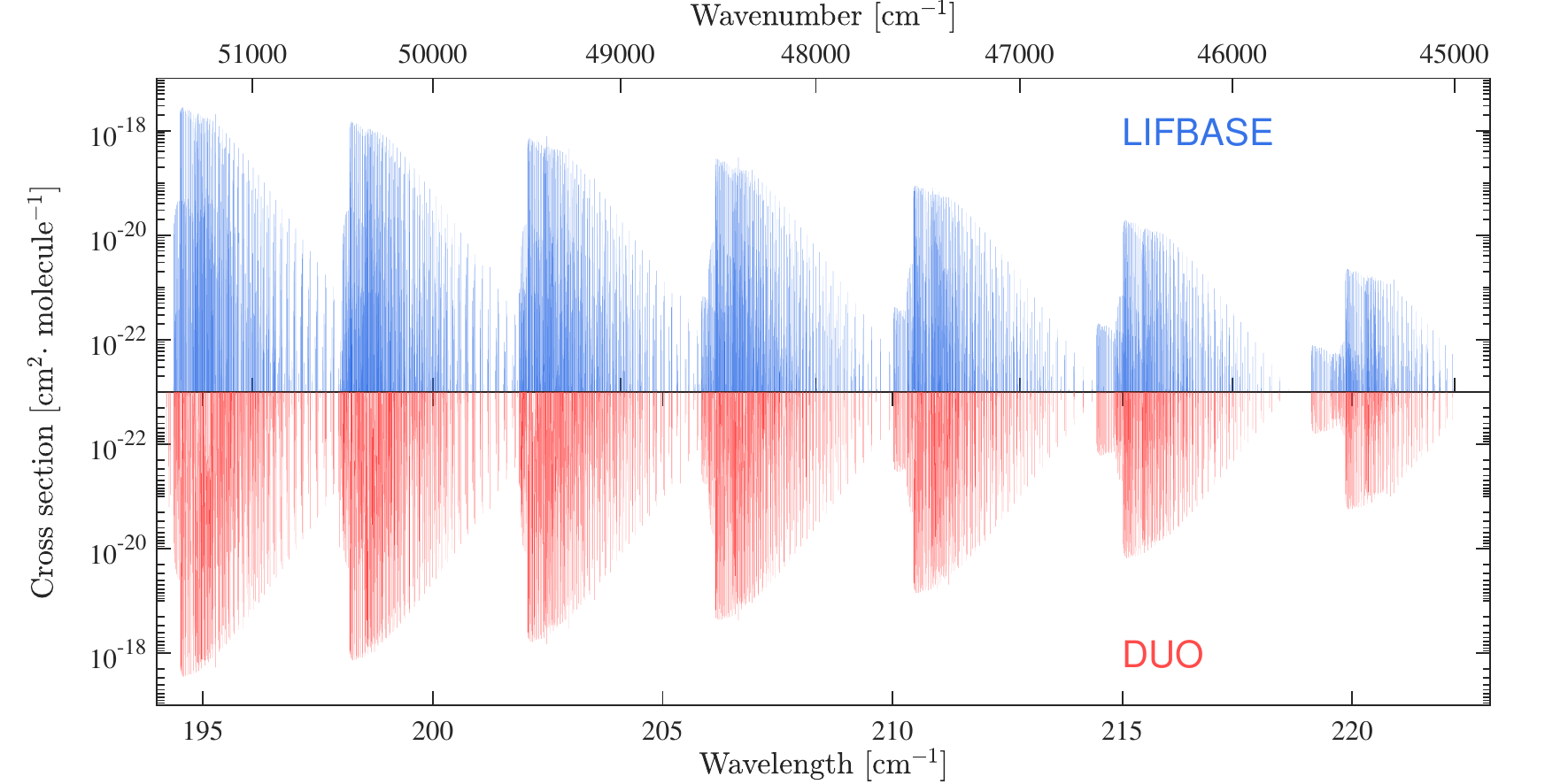}
        \caption{Calculated cross section of the NO $\upbeta(v', 0)$ ($v'=6$ to $0$ from left to right) bands of  at 295 K
            in comparison with the data from LIFBASE.
            The spectrum was computed assuming a Gaussian profile 
            with a half-width-at-half-maximum (HWHM) of \SI{0.2}{cm}.
            The relative spectrum simulated by LIFBASE is normalised to the peak 
            of $\upbeta(6, 0)$ band.}
        \label{fig:B_spectrum}
    \end{figure*}

The \BPi\,-\,\CPi  interaction have small effect on the intensities of
lower seven $\upbeta(v',0)$ bands. Figure\,\ref{fig:beta60} compares the experimental intensities of $\upbeta(6,0)$ measured by \citet{06YoThMu.NO}  and theoretical intensities calculated with \duo. 
The relative cross section values calculated by LIFBASE \citep{99LIFBASE} are compared with \duo values in Fig.\,\ref{fig:B_spectrum}.  The values of LIFBASE are scaled according to the peak of $\upbeta(6, 0)$ band.
    
\begin{figure*}
        \centering
        \includegraphics{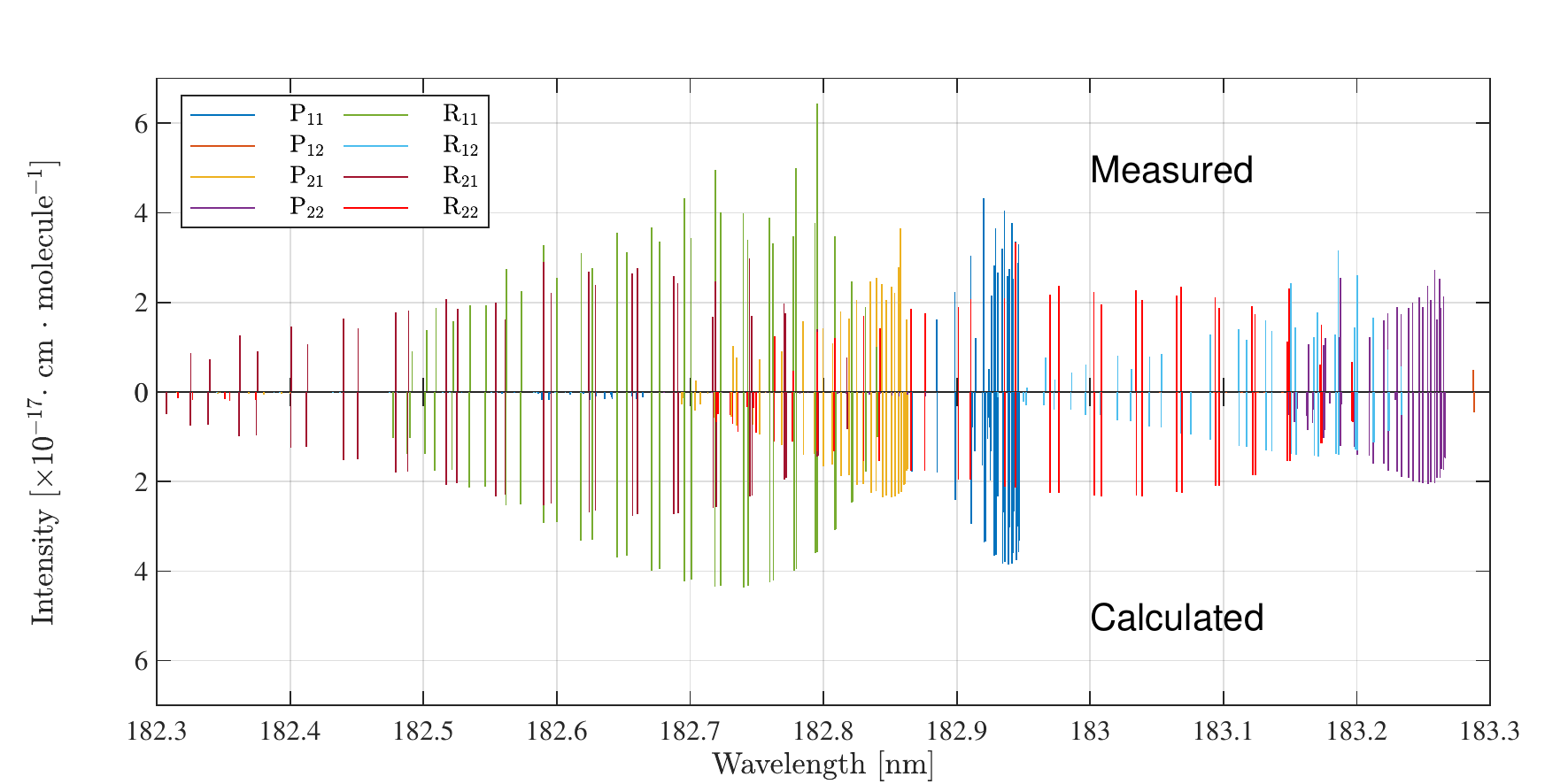}
        \caption{Calculated absorption intensities of the $\updelta(1, 0)$ band at 295 K 
            in comparison with 
            the intensities published by \citet{06YoThMu.NO}.
            This is a strong band and most of the
            $\Lambda$-doublets were resolved in the experiment.
            To allos comparisons of the fine-structure,
            we evenly divided the measured intensities 
            of any blended lines to  
            created effective $e/f$ transitions.}
        \label{fig:delta10}
    \end{figure*}

The $\updelta(1,0)$ band is the strongest one at \SI{295}{K} in the \BPi\,-- \CPi interaction region and  the intensities of the transitions in this band are plotted in Fig.\,\ref{fig:delta10}. The figure demonstrates the overall agreement between experimental and theoretical values but also exposes defects in  our model showing 
the interaction model for \BPi\,-- \CPi
is not perfect.

Note that, the  spectra in Figs. \ref{fig:gamma30}, \ref{fig:beta60} and \ref{fig:delta10}  were calculated using the pure \duo energies before they been replaced by MARVEL or EH values. However the difference between the experimental and calculated lines is  indistinguishable at this scale.

\section{Conclusions}
    
    Here we present a new line list for \NO\ called \XABC which covers transitions between the  ground electronic state \XPi and the four lowest-lying states, \XPi, \ASigma, \BPi and \CPi.  The line list combines effective Hamiltonian (\textsc{SPFIT}), \marvel and calculated (\duo) energies,  providing  high-accuracy line positions. Combined with   our \ASigma -- \XPi, \BPi -- \XPi and   \CPi -- \XPi transition dipole moments,   the diabatic model predicts the transition intensities  which agree well with the measured values. The line list is part of ExoMol project \citep{jt528}  and available from      (\href{www.exomol.com}{www.exomol.com})
    and CDS database
    (\href{http://cdsarc.u-strasbg.fr}{cdsarc.u-strasbg.fr}).

\section*{Acknowledgements}
We thank Johannes Lampel for drawing this problem to our attention and for helpful discussions. 
    Qianwei Qu acknowledges the financial support from University College London
    and China Scholarship Council.
    This work was supported by the STFC Projects No. ST/M001334/1 and ST/R000476/1, and ERC Advanced Investigator Project 883830.
    The authors acknowledge the use of the UCL Myriad, Grace and Kathleen High Performance Computing Facilities and associated support services in the completion of this work.

\section*{Data Availability}

The \duo model file and calculated partition function file are
attached as a supplementary materials.
The \XABC states and transition files of \NO can be downloaded from
\href{www.exomol.com}{www.exomol.com} and
\href{http://cdsarc.u-strasbg.fr}{cdsarc.u-strasbg.fr}. The open access  programs \exocross\ and \duo\ are  available from \href{https://github.com/exomol}{github.com/exomol}.



\bibliographystyle{mnras}
\bibliography{bib_journals_astro, bib_jtj, bib_linelists, bib_NO, bib_programs,bib_planets, bib_NO3}








\bsp	
\label{lastpage}
\end{document}